# The origin of p-type conduction in (P, N) co-doped ZnO

Ren-Yu Tian and Yu-Jun Zhao*

Department of physics, South China University of Technology,

Guangzhou 510640, P.R. China

Abstract

P mono-doped and (P, N) co-doped ZnO are investigated by the first-principles calculations. It is found that substitutive P defect forms a deep acceptor level at O site ($P_O$) and it behaves as a donor at Zn site ($P_{Zn}$), while interstitial P ($P_i$) is amphoteric. Under equilibrium conditions, these defects contribute little to the p-type conductivity of ZnO samples since the formation energy of $P_{Zn}$ is much lower than that of $P_i$ or $P_O$ when $E_F$ is below mid-gap (a prerequisite p-type condition). Zinc vacancies ($V_{Zn}$) and $P_{Zn}$-2$V_{Zn}$ complex are demonstrated to be shallow acceptors with ionization energies around 100 meV, but they are easily compensated by $P_{Zn}$ defect. Fortunately, $P_{Zn}$-4$N_O$ complexes may have lower formation energy than that of $P_{Zn}$ under Zn rich condition by proper choices of P and N sources. In addition, the neutral $P_{Zn}$-3$N_O$ passive defects may form an impurity band right above the valence band maximum of ZnO as in earlier reported (Ga,N) or (Zr,N) doped ZnO. This significantly reduces the acceptor level of $P_{Zn}$-4$N_O$ complexes, and helps improving the *p*-type conductivity in ZnO.

* zhaoyj@scut.edu.cn

## I. INTRODUCTION

Zinc oxide (ZnO) is a good candidate for applications in short-wavelength and transparent optoelectronic devices for its unique physical properties. Yet difficult realization of reproducible low resistivity *p*-type ZnO hinders its applications. There have been many attempts, such as mono-doping, [1-10] co-doping or cluster doping technology [11-14] to overcome the *p*-type doping bottleneck. Among those attempts, donor-acceptor co-doping method is promising in terms of enhancing the acceptor concentration and lowering the acceptor level. [13, 14] A record of *p*-type electrical resistivity of 0.026 Ωcm, Hall mobility of 4.4 $cm^2V^{-1}s^{-1}$, and carrier concentration of $5.5\times10^{19}$ $cm^{-3}$ has been reported with zirconium (Zr) and nitrogen (N) co-doping ZnO.[14] Reduction of the transition levels through the effective impurity bands induced by donor-acceptor complexes have successfully explained the experiment observation in Ga-N[15, 16] and Zr-N[16] co-doped *p*-type ZnO. Recently, co-doping with group V element and N, such as As-N [17] and P-N, [18] have also realized p-type ZnO. A detailed explanation of this dual-acceptor doping mechanism is not available so far, although it could be viewed as an attempt following the earlier research on group III-N and group IV-N co-doping.

In order to clarify the *p*-type conduction in (P, N) co-doped ZnO samples, P mono-doping and (P, N) co-doping in ZnO are studied by first-principles calculations in this work. By comparing the defect formation energies, we find that there are three kinds of defects, i.e., $V_{Zn}$, $P_{Zn}$-$2V_{Zn}$, and $P_{Zn}$-$4N_O$ complexes may be shallow acceptors. Furthermore, it is found that acceptors $V_{Zn}$ and $P_{Zn}$-$2V_{Zn}$ may not be responsible to the high hole carrier concentrations of about $10^{17}$~$10^{19}$ $cm^{-3}$ [3, 4] as observed experimentally due to the compensation effects of $P_{Zn}$ under equilibrium conditions, no matter what kind of P or N sources are adopted. However, $P_{Zn}$-$4N_O$ complex may play a critical role in *p*-type conductivity in (P, N) co-doped ZnO under Zn-rich conditions by using P with low $\Delta\mu_P$ and N with high $\Delta\mu_N$ dopant sources. Special attentions have also paid to the mechanism of the lowered transition level of $P_{Zn}$-$4N_O$ complex by passive impurity band.

## II. COMPUTIONAL METHODS

Our calculations are carried out with the Vienna *ab initio* simulation package (VASP), [19, 20] using the generalized-gradient approximation (GGA) formulism of PW91 [21, 22] for the exchange correlation. The electron-ion interactions are described by the projector augmented wave method (PAW). [23, 24] Zinc 3d states are treated as valence electrons. The energy cutoff for the plane wave expansion is set to 500 eV. All atoms are fully relaxed during the calculation until the Hellmann–Feynman forces converge to 10 meV/Å. The optimal crystal lattice constants of ZnO (*a*=3.28Å, *c*=5.27 Å) are adopted through the calculations, which are in excellent agreement with the experimental values (*a*=0.325 nm, *c*=0.521 nm). A 72-atom supercell is used to simulate P mono-doping and (P, N) co-doping in ZnO. The Brillouin zones are sampled with gamma centered $3\times3\times2$ *k*-point mesh. For charged defects, a uniform background charge is added to keep the global charge neutrality of the supercells.

The defect formation energy, $\Delta H_f(\alpha,q)$, for a supercell containing defect $\alpha$ in charge state $q$, can be computed as: [25-27]

$$\Delta H_f(\alpha,q) = \Delta E(\alpha,q) + \sum n_i \Delta\mu_i + qE_{\mathrm{F}} , \qquad (1)$$

where $\Delta E(\alpha,q) = E(\alpha,q) - E(\mathrm{host}) + \sum n_i \mu_i(\mathrm{bulk}) + q\varepsilon_{\mathrm{VBM}}(\mathrm{host})$. $E(\alpha,q)$ is the total energy for the studied supercell containing defect $\alpha$ in charge state $q$ and $E(\mathrm{host})$ is the total energy of the same supercell without the defect. $n_i$ is the number of atom $i$ involved in the defect, and $q$ is the number of electrons transferred from the supercell to the reservoirs in forming the defect cell. $E_{\mathrm{F}}$ is the electron Fermi energy referenced to the valence-band maximum (VBM) of the host, and varies up to the experimental bang gap value of 3.37 eV. [28] $\varepsilon_{\mathrm{VBM}}$ is the VBM energy of the host supercell. $\Delta\mu_i$ is the chemical potential of atom $i$ referenced to its elemental solid/gas with cohesive energy of $\mu_i(\mathrm{bulk})$, and should not be greater than 0 eV generally in order to avoid precipitation of the elemental solid/gas. Meanwhile, to maintain a stable compound ZnO and to avoid other possible competing phases, it must satisfy the following conditions

$$\Delta\mu_{\mathrm{Zn}} + \Delta\mu_{\mathrm{O}} = \Delta H_f(\mathrm{ZnO})$$

$$j \cdot \Delta\mu_{\mathrm{Zn}} + k \cdot \Delta\mu_{\mathrm{O}} + m \cdot \Delta\mu_{\mathrm{P}} + n \cdot \Delta\mu_{\mathrm{N}} \leq \Delta H_f(\mathrm{Zn}_j\mathrm{O}_k\mathrm{P}_m\mathrm{N}_n) \qquad (2)$$

Where $\mathrm{Zn}_j\mathrm{O}_k\mathrm{P}_m\mathrm{N}_n$ stands for possible alloys formed by Zn, O, P and/or N naturally except for ZnO. $\Delta H_f(\mathrm{ZnO})$ and $\Delta H_f(\mathrm{Zn}_j\mathrm{O}_k\mathrm{P}_m\mathrm{N}_n)$ represent the formation energies of ZnO and the corresponding competing phases. The calculated cohesive energies of elements and the calculated formation energies of possible competing compounds in (P, N) co-doped ZnO are listed in Table 1.

The defect transition energy level $\varepsilon_{\alpha}(q/q')$ is the $E_{\mathrm{F}}$ in Eq.(1), at which the formation energy $\Delta H_f(\alpha,q)$ of defect $\alpha$ in charge state $q$ is equal to that of another charge $q$' of the same defect, i.e.,

$$\varepsilon_{\alpha}(q/q') = [\Delta E(\alpha,q) - \Delta E(\alpha,q')]/(q' - q) \tag{3}$$

In this paper, we use the hybrid scheme combining *k*-points sampling and Γ-point-only approaches to calculate the transition energy level and the defect formation energy.[27, 29] In this scheme, the transition energy level for acceptors ($q$<0) with respect to VBM is given by:

$$\varepsilon(0/q) = [\varepsilon_D^{\Gamma}(0) - \varepsilon_{\mathrm{VBM}}^{\Gamma}(\mathrm{host})] + [E(\alpha,q) - (E(\alpha,0) - q\varepsilon_D^{k}(0))]/(-q) \tag{4}$$

For donors ($q$>0), the ionization energy referenced to the conduction-band minimum (CBM) is given by:

$$\varepsilon_g^{\Gamma}(\mathrm{host}) - \varepsilon(0/q) = [\varepsilon_{\mathrm{CBM}}^{\Gamma}(\mathrm{host}) - \varepsilon_D^{\Gamma}(0)] + [E(\alpha,q) - (E(\alpha,0) - q\varepsilon_D^{k}(0))]/q \tag{5}$$

where $\varepsilon_D^{k}(0)$ and $\varepsilon_D^{\Gamma}(0)$ are the defect levels at the special *k*-points (averaged) and at the Γ-point, respectively; $\varepsilon_{\mathrm{VBM}}^{\Gamma}(\mathrm{host})$ and $\varepsilon_{\mathrm{CBM}}^{\Gamma}(\mathrm{host})$ are the VBM and CBM energy, respectively, of the host at the Γ-point; and $\varepsilon_g^{\Gamma}(\mathrm{host})$ is the calculated bandgap at the Γ-point. Meanwhile, the average electrostatic potential at core area of oxygen atom far away from the defect is adopted as a reference to determine the VBM alignment.

The formation energy of a charged defect is then given by:

$$\Delta H_f(\alpha,q) = \Delta H_f(\alpha,0) - q\varepsilon(0/q) + qE_{\mathrm{F}} \tag{6}$$

where $\Delta H_f(\alpha,0)$ is the formation energy of the charge-neutral defect.

Although a 72-atom supercell is adopted in the calculations, the defect concentration of the studied system is around $10^{21}$-$10^{22}$ $cm^{-3}$, which is significantly higher than a typical doping concentration. The fictitious high doping concentration may imply unwanted impurity interactions, especially for the charged defects. Here a band filling correction is employed to get a more reasonable transition level following the approach described in Ref. 26. To be consistent with earlier theoretical work on p-type doping in ZnO, the image charge correction is not employed in this work. In fact, we have checked the transition level of $As_{Zn}$-$2V_{Zn}$ to validate our approach on the transition level calculation. Our calculated transition level $\varepsilon(0/-)$ is 0.14eV for $As_{Zn}$-$2V_{Zn}$, which is in good agreement with earlier reported value of 0.15eV by Limpijumnong et al. [30]

**III. Monodoping of P in ZnO**

Figure 1 shows formation energies of defects related to phosphorous mono-doping in ZnO under the two extreme conditions: (a) zinc-rich limit, i.e., $\Delta\mu_{\mathrm{Zn}} = 0$ and $\Delta\mu_{\mathrm{O}} = \Delta H_f(\mathrm{ZnO})$ ; and (b) oxygen-rich limit, i.e., $\Delta\mu_{\mathrm{O}} = 0$ and $\Delta\mu_{\mathrm{Zn}} = \Delta H_f(\mathrm{ZnO})$. Here we suppose $P_2O_5$ is adopted as the P source. The slope corresponds to the charge state $q$ as used in Eq.(1)~Eq.(6). A change in the slope indicates transition of the charge state. The transition levels are independent of the choice of atomic chemical potentials and thus they are the same in Fig.1(a) and Fig.1(b). Similar to the reported P[31] and As[30] dopants, interstitial P ($P_i$) is amphoteric as diluted point defects. A substitutional P at an O lattice site ($P_O$) forms a deep acceptor level of $\varepsilon(0/-) = 0.96$ eV, which is in excellent agreement with earlier calculation of 0.93 eV. [32] A substitutional P at a Zn lattice site ($P_{Zn}$) behaves as a donor. The formation energy of $P_{Zn}$ is much lower than that of $P_i$ and $P_O$ in most situations except that the Fermi level is greater than 2.98 eV above the VBM under O-rich condition. This means that none of these three point defects could contribute to the p-type conductivity remarkably in ZnO.

Recent calculations have argued that in ZnO doped with $P_2O_5$, the dominant acceptors are $V_{Zn}$ defects, while the $P_{Zn}$-$2V_{Zn}$ complex is energetically more favorable for the $Zn_3P_2$ source under O-rich conditions. [31] Yet experiment results indicated that P atoms replace the O atoms to form $P_O$ rather than forming $P_{Zn}$-$2V_{Zn}$ complex as acceptors. [1] Here, the formation energy of $P_{Zn}$-$2V_{Zn}$ complex and $V_{Zn}$ are also plotted in Fig.1 for comparison. Figure 2 indicates the most stable structure of $P_{Zn}$-$2V_{Zn}$ complex, which is consistent with that reported by Lee et.al. [31] Contrary to the reported deep transition level (0.55 eV above the VBM), [31] we find that $P_{Zn}$-$2V_{Zn}$ complex may be a shallow acceptor defect, with the ionization energy of $\varepsilon(0/-) = 0.12$ eV. On the other hand, the formation energy of $P_{Zn}$-$2V_{Zn}$ complex is 4.62 eV (Zn-rich) and 2.34 eV (O-rich) with $P_2O_5$ source for P source, which is much greater than that of $P_{Zn}$ when the Fermi level is close to VBM. This indicates that $P_{Zn}$-$2V_{Zn}$ complex may be fully compensated by $P_{Zn}$ when $P_2O_5$ source is adopted. The calculated transition levels of $V_{Zn}$ are $\varepsilon(0/-) = 0.091$ eV and $\varepsilon(-/2-) = 0.42$ eV, indicating that $V_{Zn}$ may act as a shallow acceptor. However, its formation energy is relatively high compared with that of $P_{Zn}$ under both Zn-rich and O-rich conditions when $E_F$ is near VBM. It also suffers the compensating effect of $P_{Zn}$, thus unlikely to play important role in *p*-type conductivity, which requires the Fermi level of ZnO below the mid gap.

**IV. (P, N) co-doping in ZnO**

Recently, it was proposed that the introduction of mutually passivated impurity bands may successfully overcome the doping asymmetry in ZnO by (Ga, N)[15, 16] or (Zr, N)[16] co-doping. Consequently, passive defect complexes might be formed in group V, N (such as P, N) co-doped ZnO. Actaully, there are two noticeable configurations of (P-3N) complex (as shown in Fig. 3) when P substituting Zn and 3 N replacing three nearest neighbor O atoms. One is the three nearest in-plane O atoms replaced by N atoms [Fig.3(a)], and the other is N atoms substituting the nearest out-of-plane O atom together with two of the nearest in-plane O atoms[Fig.3(b)]. The configuration shown in Fig.3(a) is considered as the passive stoichiometric (P-3N) complex, since it is energetically favored by 0.039 eV than the latter one. The calculated total DOS for the pure ZnO host and the passive stoichiometric (P-3N) complex doped ZnO are shown in Fig. 4. Unlike the additional fully occupied impurity band in (Ga, N)[15, 16] and (Zr, N)[16] co-doped ZnO, the valence band appears to be modified by the passive stoichiometric (P-3N) complex, with its VBM shifted up by 0.156eV with respect to that of the pure ZnO host. We have carefully checked the partial DOS of the passive stoichiometric complexes in (Ga, N) and (P, N) co-doped ZnO to investigate the property of the modified valence band in (P, N) co-doped ZnO, and found that the electronic states near the VBM of the doped systems is mainly constituted of the *p* states of N dopants and *d* states of Zn. The *p* states of N dopants of the passive stoichiometric complexes in (Ga, N) and (P, N) co-doped ZnO, as well as the *d* states of Zn of the two doped systems and the ZnO host are compared in Fig. 5. It can be clearly seen that the upper part of the valence band in (P, N) co-doped ZnO has the same origin with (Ga, N) co-doped ZnO, i.e., the hybridization of the *p* states of N dopants and *d* states of Zn of ZnO host. In other words, the passive stoichiometric (P-3N) complex also forms an additional fully occupied impurity band, although it is not obviously shown in DOS plot as that in (Ga, N)[15, 16] and (Zr, N)[16] co-doped ZnO. With the impurity band above VBM of pure ZnO, the acceptor level is expected to decrease when additional N atoms are introduced into the (P-3N) co-doped ZnO since electrons may transit from the impurity band.

When an additional N is introduced to the above mentioned two configurations of (P-3N) complex, we have investigated nine unequal configurations of P-4N complex for the additional N atom replacing the nearest and the second nearest neighbor O site of $P_{Zn}$. We find that the configuration with four N atoms substituting the first neighbor O sites of the $P_{Zn}$ is energetically favored, while the other configurations are at least 0.64 eV higher in total energy. This clearly indicates that the doped P and N tend to form $P_{Zn}$-$4N_O$ complexes in ZnO.

For the dopant source of $P_2O_5$ and $N_2$, the calculated formation energy of the $P_{Zn}$-$4N_O$ complex is shown in Fig. 6. The ionization energy of $P_{Zn}$-$4N_O$ is 0.271 eV above the VBM of ZnO host according to Eq.(4), yet it is reduced to 0.115 eV from the impurity band of the passive ($P_{Zn}$-$3N_O$) complex. Fig.6 also shows that the formation of $P_{Zn}$-$4N_O$ in Zn-rich condition is -1.28 eV, which is 18.87 eV lower than that of O-rich condition, indicating that *p*-type conductivity is easier to achieve in Zn-rich condition when $P_2O_5$ and $N_2$ are used as dopants in ZnO simultaneously.

The formation energies of three kinds of accepters with low transitional level, $P_{Zn}$-$4N_O$, $P_{Zn}$-$2V_{Zn}$ and $V_{Zn}$, under two extreme conditions, (a) for Zn-rich and (b) for O-rich, are plotted in Fig. 7. The most possible killer of acceptors, $P_{Zn}$, is also shown in Fig.7. We can see that for the three acceptors, $P_{Zn}$-$4N_O$ is energetically favorable under Zn-rich condition; while $V_{Zn}$ is favored under O-rich condition when $P_2O_5$ and $N_2$ are served as dopant sources. Yet the concentration of hole carriers is expected not to be much for the compensation effect of $P_{Zn}$ donor in both situations for $P_2O_5$ and $N_2$ sources (Solid line in Fig.7).

We notice that it is rare to conduct the (P, N) codoping under P-rich and N-rich condition, i.e, the chemical potentials of P and N are in equilibrium with those of $P_2O_5$ and $N_2$, respectively. Additionally, most experimentally reported *p*-type ZnO samples are obtained with NO source instead of $N_2$. Therefore, there are rooms to further investigate the *p*-type doping possibility of (P, N) by adjusting the chemical potential of the dopants. [33] Since the change of $\Delta\mu_P$ will simultaneously shift the formation energies of $P_{Zn}$-$2V_{Zn}$ and $P_{Zn}$, the adjustment of the chemical potentials of the dopants (i.e., P and N) has no influence to the relative formation energy of $P_{Zn}$-$2V_{Zn}$ to $P_{Zn}$ [Fig. 7(b), dashed line]. This means that although $P_{Zn}$-$2V_{Zn}$ complex may energetically more favorable than $V_{Zn}$, it may not be accounted for the *p*-type conductivity in ZnO merely through adjusting the chemical potential of P and/or N under O-rich condition. In other words, adjustment of $\Delta\mu_{\rm P}$ and $\Delta\mu_{\rm N}$ has no impact of pinned Fermi level [Fig.7(b)] under O-rich condition.

Considering the formation energy difference between $V_{Zn}$ and $P_{Zn}$ ($q=0$) is about 6.30 eV [solid line in Fig.7(b)], it is very difficult to overcome the energy difference by adjusting $\Delta\mu_{\rm P}$ experimentally. However, the situation may be changed under Zn-rich condition. When the chemical potential of P source is reduced and that of N source is raised (e.g., by using NO, $NO_2$ sources), the formation energy of $P_{Zn}$-$4N_O$ will decrease while that of $P_{Zn}$ will increase. This will enhance the *p*-type conductivity in ZnO. For example, when the chemical potential of P source is reduced to $\Delta\mu_{\rm P}=-3.0$ eV, and that of N source raised to $\Delta\mu_{\rm N}=1.7$ eV, the pinned Fermi level will decrease to around the transition level of $P_{Zn}$-$4N_O$ [ shown with dashed line in Fig.7(a)], then the compensation effect of $P_{Zn}$ becomes insignificant and *p*-type conductivity in ZnO will be greatly enhanced. The formation energy of $P_{Zn}$-$4N_O$ could even be tuned lower than that of the compensator by further adjustment of $\Delta\mu_{\rm P}$ and $\Delta\mu_{\rm N}$ when necessary. From this point of view, we tentatively suggest preparing (P, N) co-doped *p*-type ZnO under Zn-rich condition, rather than under O-rich condition.

**IV. SUMMARY**

We have investigated the formation of isolated defects and defect complexes in P-mono doped and (P, N) co-doped ZnO samples through the first-principles calculations. Although $V_{Zn}$ and $P_{Zn}$-$2V_{Zn}$ complex show low ionization energy, they pay little contribution to *p*-type conductivity of samples for the strong compensation effects of $P_{Zn}$ donors. The $P_{Zn}$-$3N_O$ passive defects may form an impurity band right above the VBM due to the hybridization of the *p* states of N and the *d* states of ZnO, as in (Ga,N) and (Zr,N) doped ZnO systems. Thus the ionization energy of $P_{Zn}$-4$N_O$ complex is reduced from 0.271 eV to 0.115 eV when electrons are transited from the top of impurity band. We also find that $P_{Zn}$-4$N_O$ complex may overcome the compensation effect of $P_{Zn}$ under Zn-rich condition with a proper choice of P, N dopant sources, and attribute to the reported *p*-type conductivity in (P,N) co-doped ZnO system.

## ACKNOWLEDGMENTS

We are grateful for the computer time at the High Performance Computer Center of the Shenzhen Institute of Advanced Technology (SIAT), Chinese Academy of Science.

## Figure captions

Figure 1 (Color online) The formation energies of possible P mono doped defects, $P_{Zn}$, $P_i$ and $P_O$, together with $P_{Zn}$-$2V_{Zn}$ complex and $V_{Zn}$ as a function of the Fermi energy under Zn-rich (a) and O-rich (b) condition for $P_2O_5$ source.

Figure 2 (color online) Relaxed structure of the most stable $P_{Zn}$-$2V_{Zn}$ complex

Figure 3 (Color online) Structures of (P-3N) complex when P substituting Zn and 3 N replacing three nearest neighbor O atoms. Configuration (a) denotes the three nearest in-plane O atoms replaced by N atoms, and (b) N atoms substituting the nearest out-of-plane O atom together with two of the nearest in-plane O atoms.

Figure 4 (Color online) Calculated total DOS of the pure ZnO host and (P-3N) co-doped ZnO

Figure 5 (Color online) Comparison of the *p* states of N dopants of the passive stoichiometric complexes in (Ga, N) and (P, N) co-doped ZnO (a) and *d* states of Zn of the two doped systems and the ZnO host (b).

Figure 6 The formation energy of $P_{Zn}$-$4N_O$ complex when $P_2O_5$ and $N_2$ are used as dopants. The vertical dashed line stands for the impurity band due to the passive ($P_{Zn}$-$3N_O$) complex.

Figure 7 (Color online) The formation energies of $P_{Zn}$-$4N_O$ (blue), $P_{Zn}$-$2V_{Zn}$ (red) and $V_{Zn}$ (magenta) under two extreme conditions, (a) for Zn-rich and (b) for O-rich, together with the most possible killer of acceptor, $P_{Zn}$ (Green). The solid lines indicate the formation energies for $P_2O_5$ and $N_2$ sources, while the dashed lines indicate the formation energies with $\Delta\mu_{\mathrm{P}} = -3.0\ \mathrm{eV}$ and $\Delta\mu_{\mathrm{N}} = 1.7\ \mathrm{eV}$. The vertical short dashed line stands for the additional impurity band due to the passive ($P_{Zn}$-$3N_O$) complex.

## Tables

Table 1 The cohesive energies of elements and the formation energy of possible competing compounds in (P, N) co-doped ZnO, as well as their space group

| Element | Space Group | Cohesive energy (eV) |
|---|---|---|
| Zn | P63/mmz | 1.123 |
| $O(O_2)$ | P1 | 2.889 |
| $N(N_2)$ | P1 | 5.328 |
| P | CMCA | 3.542 |
| Compound | | Formation energy (eV) |
| ZnO | P63MC | -3.525 |
| $ZnO_2$ | P21/A-3 | -3.435 |
| $NO_2$ | P1 | -0.959 |
| NO | P1 | 0.724 |
| $N_2O$ | P1 | 4.211 |
| $N_2O_3$ | P212121 | -1.974 |
| $N_2O_4$ | I2/M-3 | -3.681 |
| $N_2O_5$ | P63/mmc | -3.752 |
| $Zn_3P_2$ | P42/nmcs | -1.155 |
| $Zn_3N_2$ | I21/A-3 | 0.415 |
| $Zn_3(PO_4)_2$ | C12/C1 | -30.092 |
| $Zn_2P_2O_7$ | PBCM | -26.089 |
| $ZnP_2O_6$ | C12/C2 | -21.676 |
| $P_2O_5$ | FDD2 | -16.580 |
| $P_4O_6$ | P121/M1 | -19.334 |
| $P_3N_5$ | C12/C2 | -2.641 |

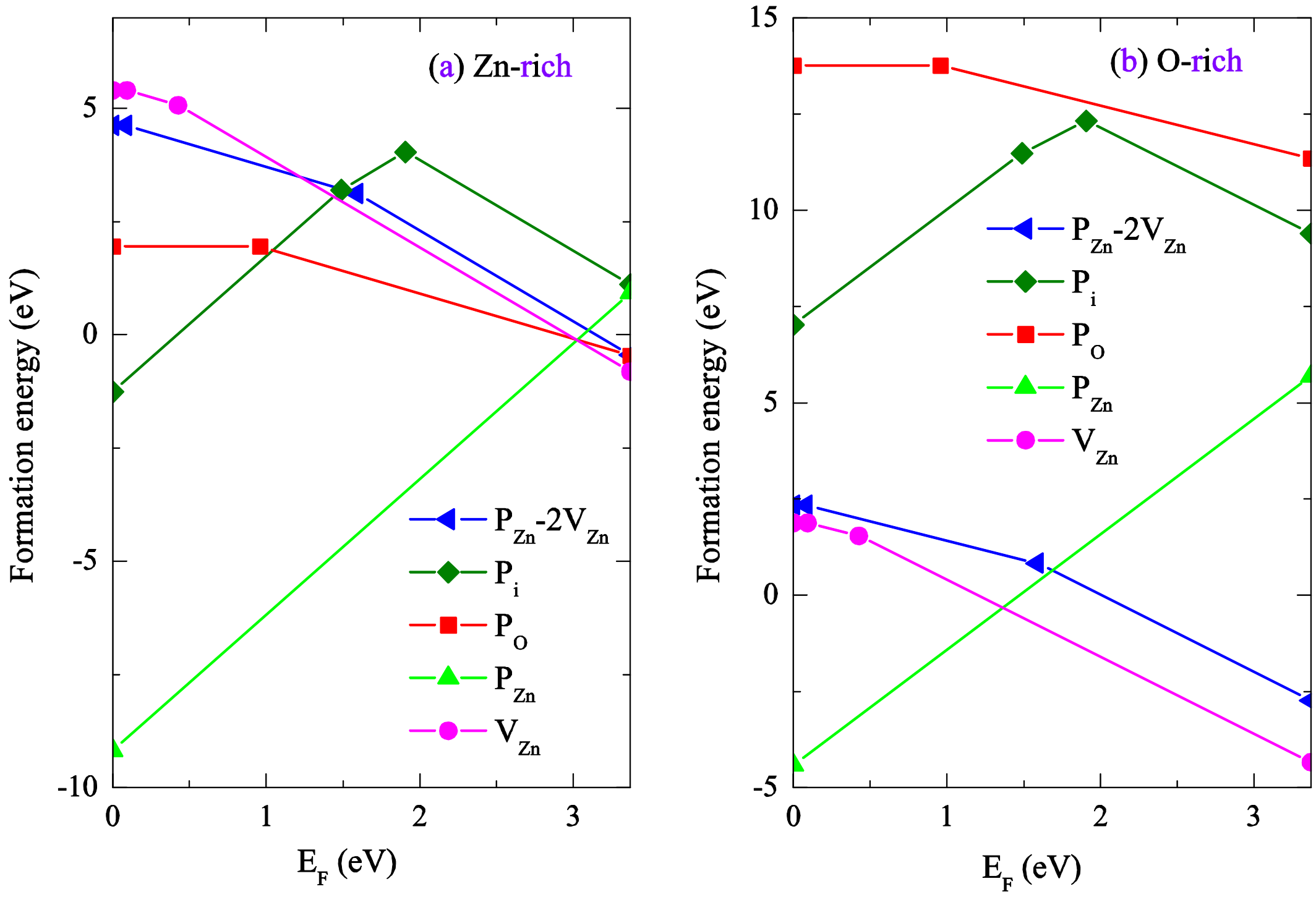

Fig. 1    Ren-Yu Tian and Yu-Jun Zhao

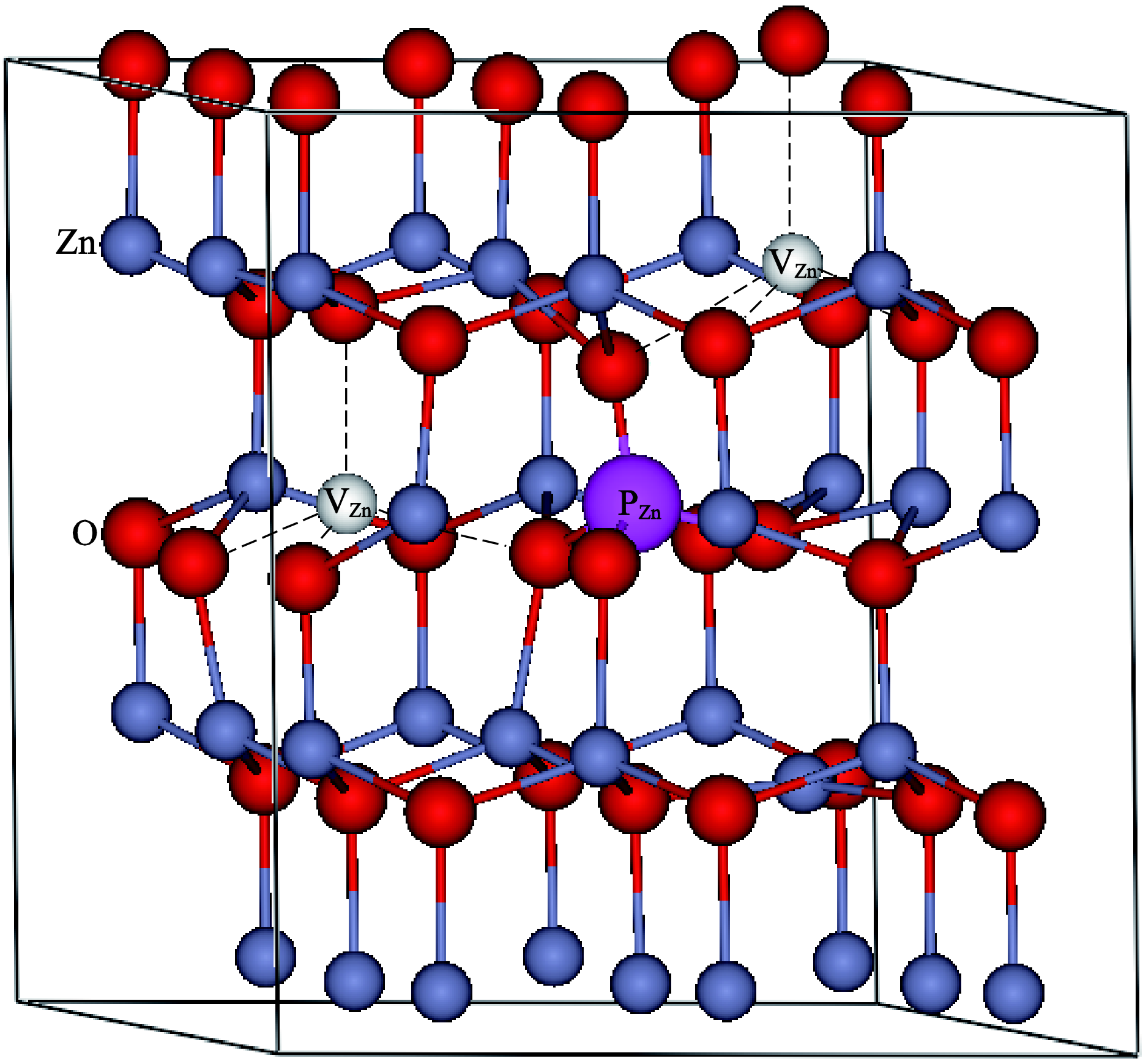

Fig. 2 Ren-Yu Tian and Yu-Jun Zhao

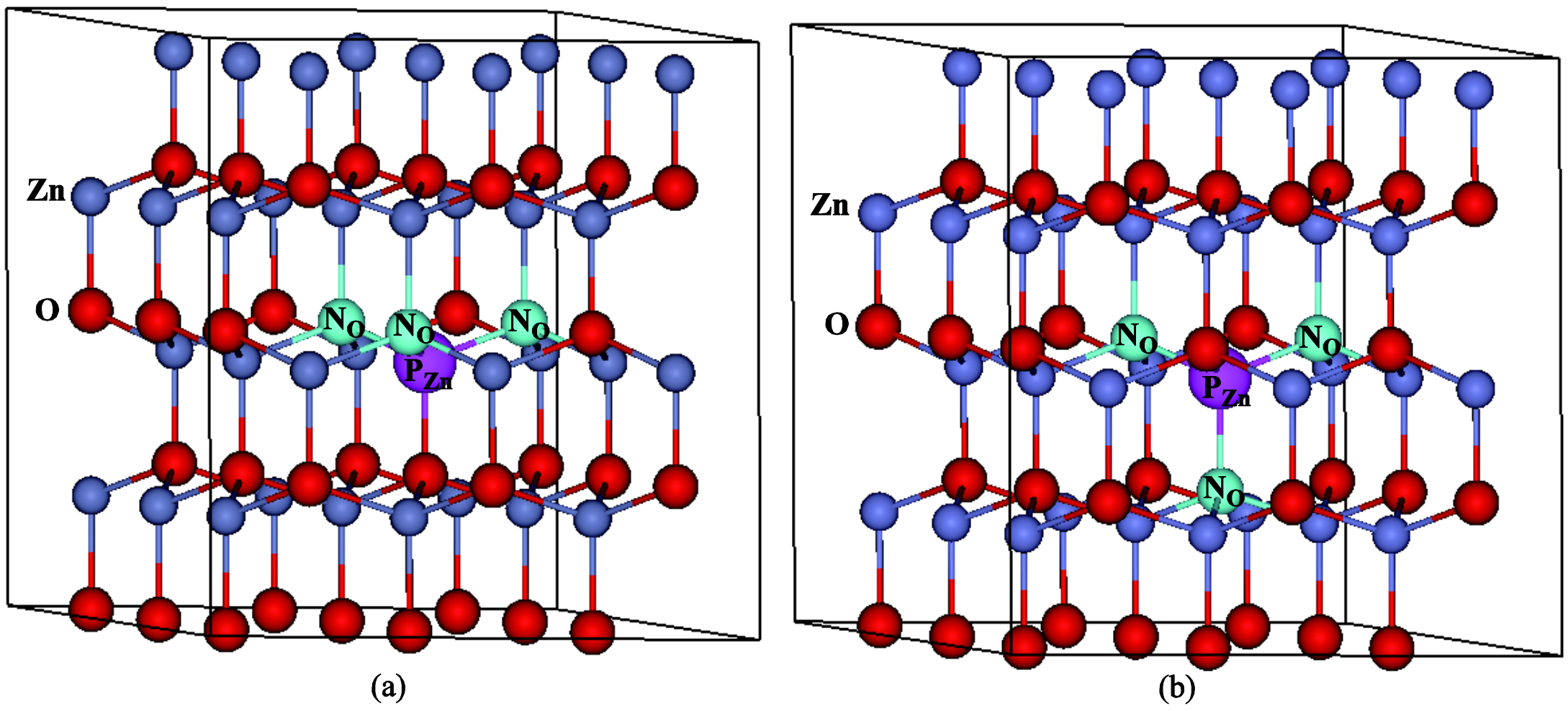

Fig. 3 Ren-Yu Tian and Yu-Jun Zhao

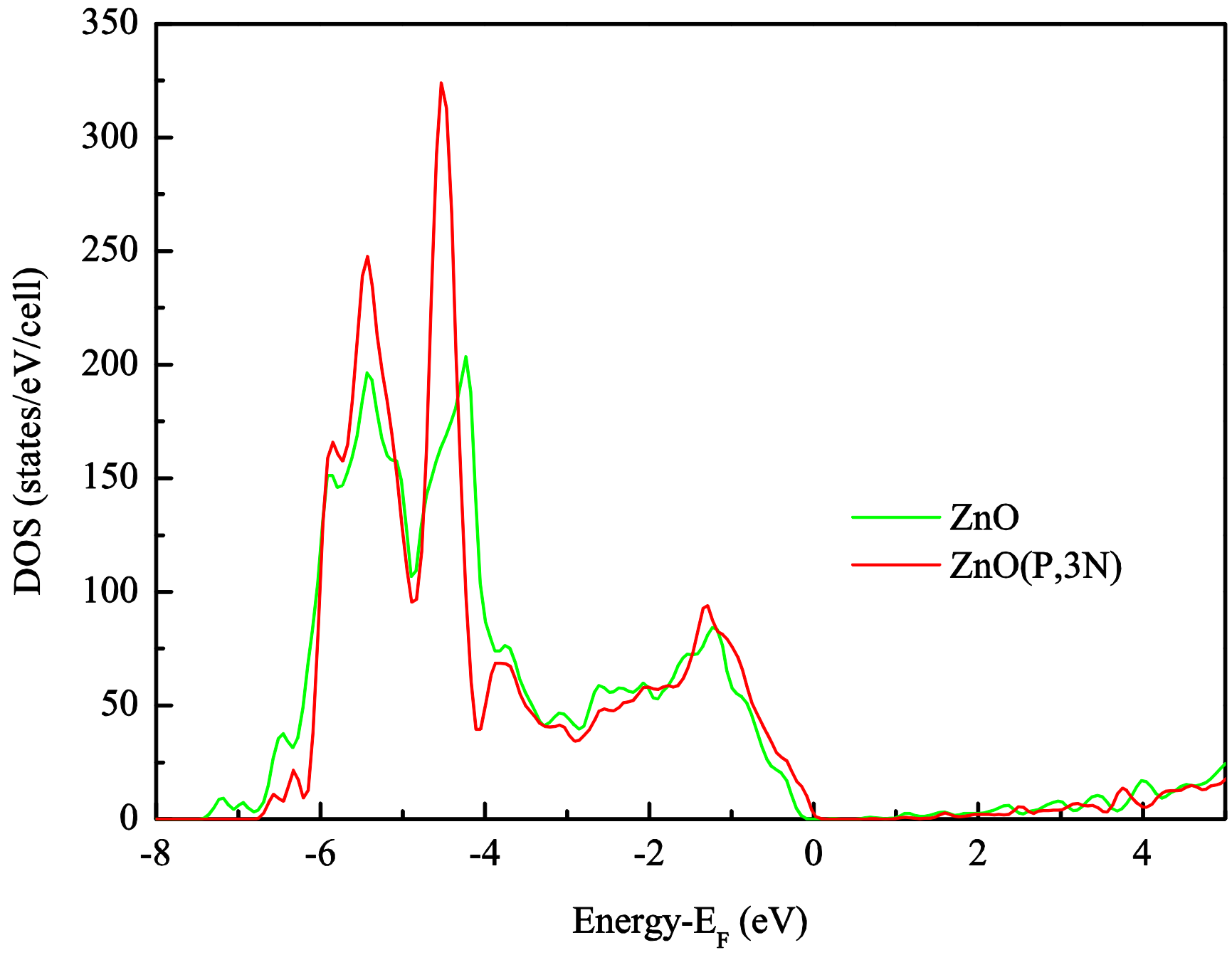

Fig. 4 Ren-Yu Tian and Yu-Jun Zhao

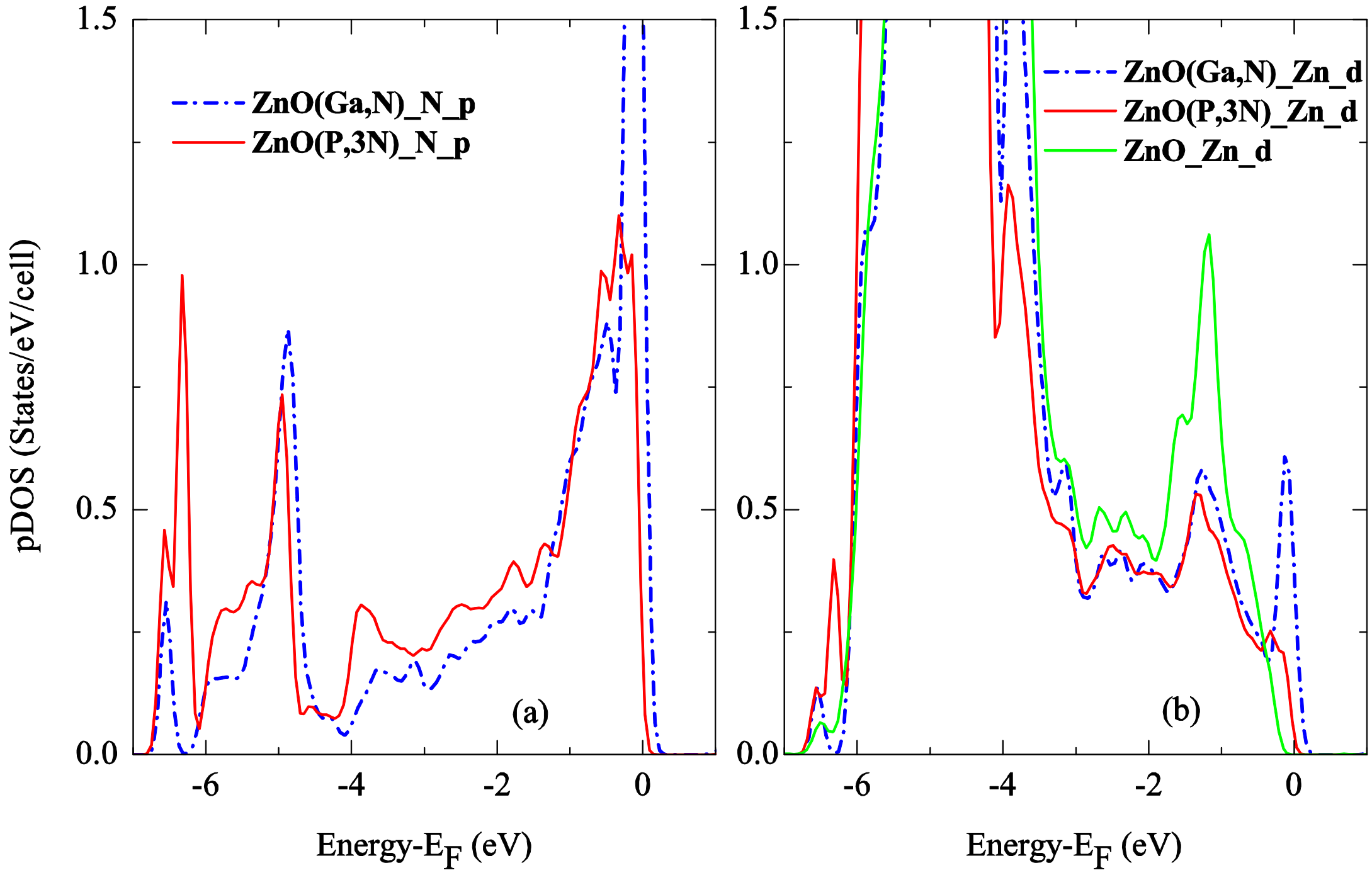

Fig. 5 Ren-Yu Tian and Yu-Jun Zhao

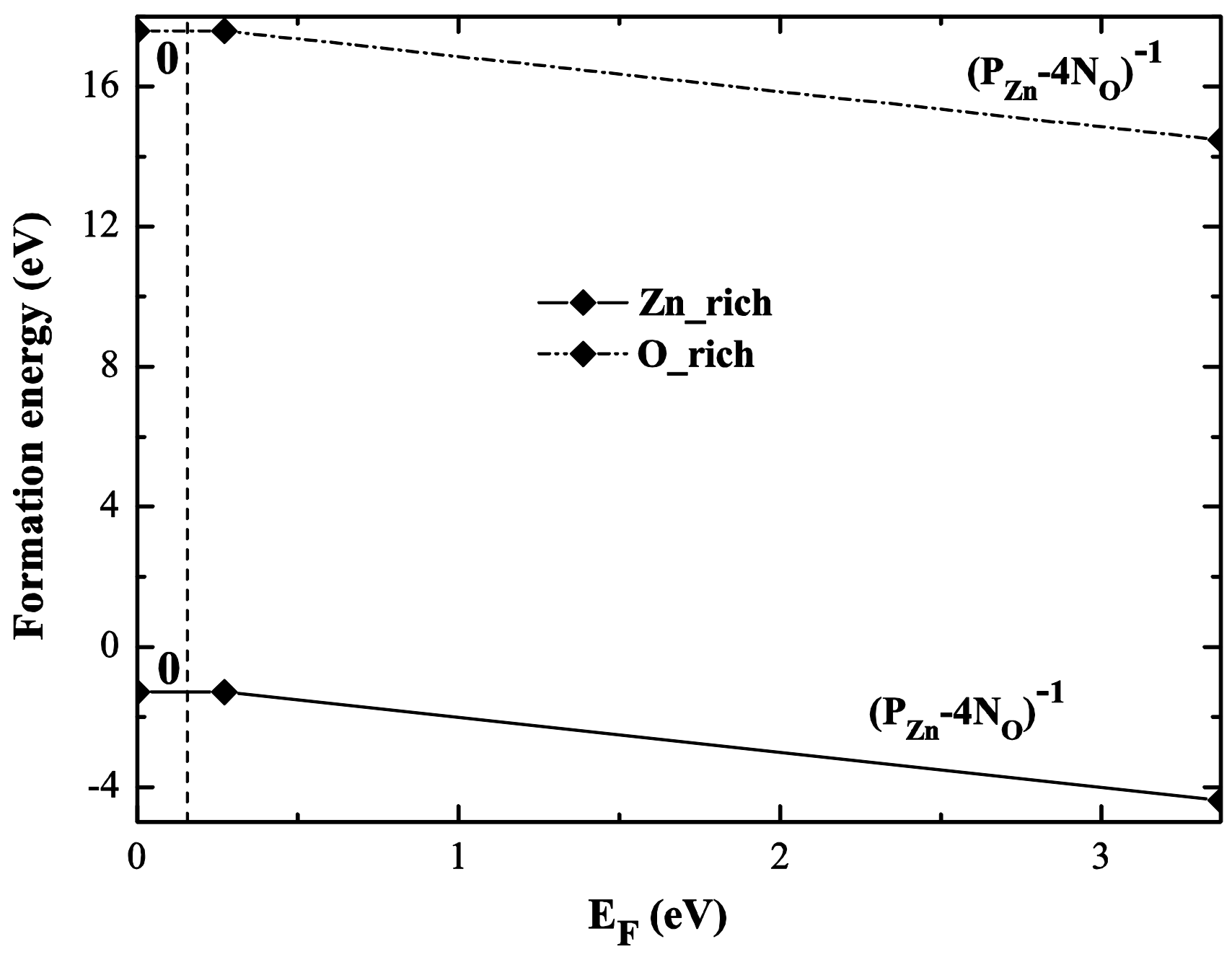

Fig. 6 Ren-Yu Tian and Yu-Jun Zhao

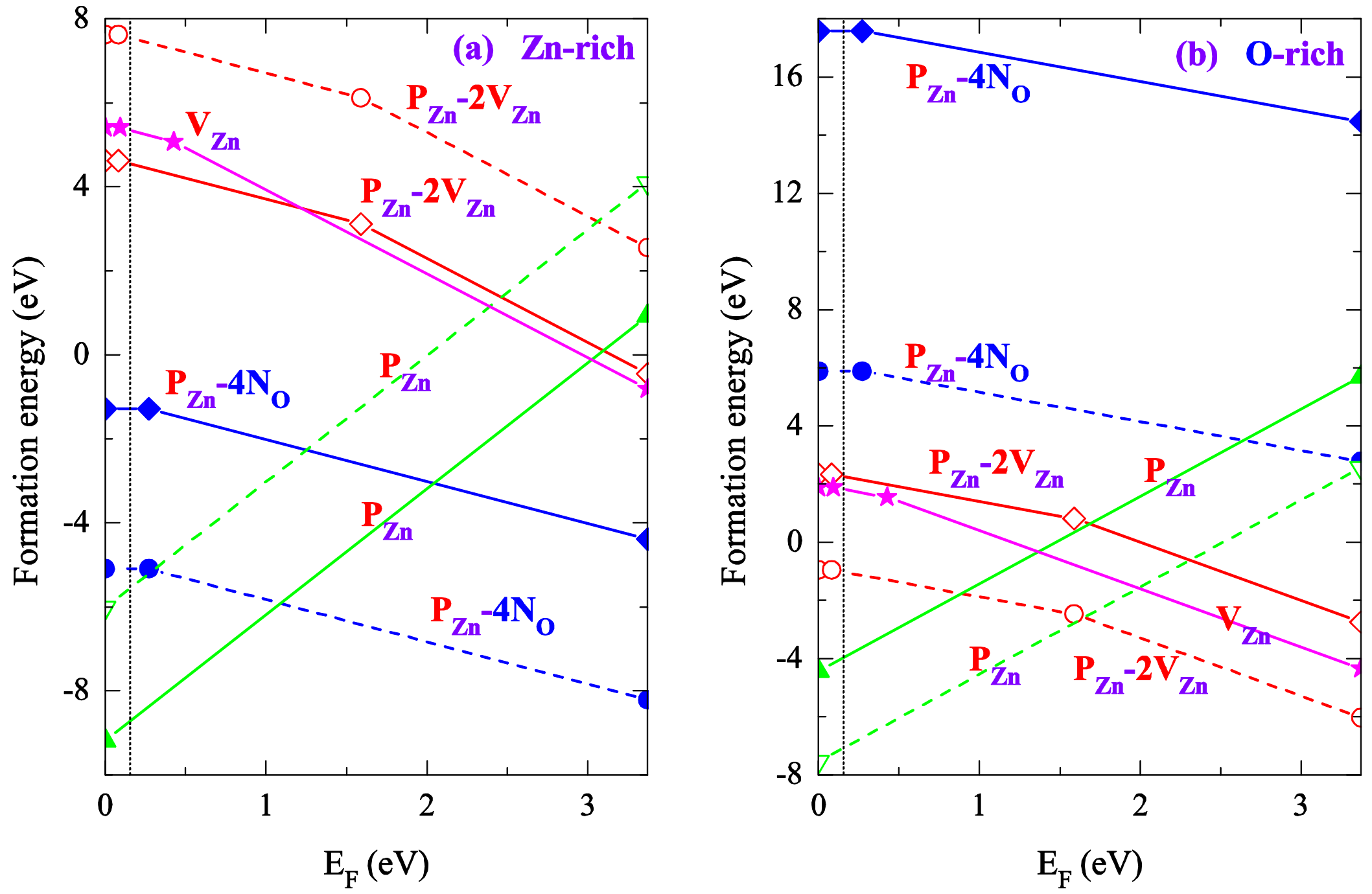

Fig. 7    Ren-Yu Tian and Yu-Jun Zhao